\documentstyle[prl,aps,epsf,floats]{revtex}
\begin{document}
\draft
\twocolumn[\hsize\textwidth\columnwidth\hsize\csname @twocolumnfalse\endcsname
\title { Ground-state dispersion and density of states from path-integral 
         Monte Carlo. \\   
         Application to the lattice polaron } 

\author{P.\,E.\,Kornilovitch}
\address{Blackett Laboratory, Imperial College,
Prince Consort Road, London SW7 2BZ, United Kingdom}

\date{\today}
\maketitle

\begin{abstract}

A formula is derived that relates the ground-state dispersion of a 
many-body system with the end-to-end distribution of paths with open 
boundary conditions in imaginary time. The formula does not involve
the energy estimator. It allows direct measurement of the ground-state
dispersion by quantum Monte Carlo methods without analytical
continuation or auxiliary fitting. The formula is applied to the lattice 
polaron problem. The exact polaron spectrum and density of states are 
calculated for several models in one, two, and three dimensions.
In the adiabatic regime of the Holstein model, the polaron density
of states deviates spectacularly from the free-particle shape.

\end{abstract}
\vskip2pc]
\narrowtext

\section{Introduction}
\label{sec:one}

Usually, quantum Monte Carlo (QMC) methods are used to study 
ground-state, thermodynamic, or static properties of quantum-mechanical 
systems. Some important dynamical characteristics can also be 
obtained through various forms of fluctuation-dissipation relations.
Examples of these are the superfluid fraction of Bose-liquids
\cite{Ceperley}, the Drude weight of conductors \cite{Scalapino},
the Meissner fraction of superconductors \cite{Scalapino}, and
the effective mass of defects \cite{Basile,Ceperley_two} and polarons
\cite{Alexandrou,Kornilovitch}. Beyond that dynamical calculations
are less straightforward. For instance, calculation of the excitation
spectrum normally requires measurement of the Green's function
at imaginary times and subsequent analytical continuation to real
times. 

However, there exists one special type of excitation spectrum that 
can be measured directly by QMC. This is the ground-state dispersion,
i.e., the total energy of the system $E_{\bf P}$ as a function of
the total momentum {\bf P}. In a translationally-invariant system,
{\bf P} is a constant of motion, and the Hamiltonian does not mix
subspaces with different {\bf P}. Then, if a QMC is designed as to
operate within a given {\bf P}-subspace only, it may be able to
access the ground state for the given {\bf P}, thereby providing
$E_{\bf P}$. Not for any physical system $E_{\bf P}$ is of interest. 
For a collection of identical particles, for instance, one has simply 
$E_{\bf P} = {\bf P}^2/(2M)$, $M$ being the total mass, which 
corresponds to free movement of the system as a whole.
Positive examples include cases when the system can be divided into a 
tagged particle and an environment (usually bosons). The best
known example of this kind is the polaron, i.e., an electron strongly
interacting with phonons. In this case $E_{\bf P}$ is nothing but
the polaron spectrum. The polaron spectrum will be the main subject
of this paper.

There exist at least two different strategies of how to operate
within a restricted {\bf P}-subspace. The first one is to work in
momentum space and to fix the total momentum of the system from
outset. An example of this approach is the diagrammatic method of 
Prokof'ev and Svistunov \cite{Prokof'ev}. In this method QMC is used 
to sum the entire diagrammatic series for an imaginary-time 
Green's function $G({\bf P}, \tau)$. Since the total momentum is an 
external parameter of the series, it is possible to extract $E_{\bf P}$ 
from the $\tau \rightarrow \infty$ limit behavior of $G$, by fitting
it to a single exponential $e^{-E_{\bf P} \tau}$. This method is exact 
and universal but requires a separate simulation for each {\bf P}-point.

The second strategy is to work in real space but to use Fourier-type
projection operators to project on states with definite {\bf P}.
This amounts to the free boundary conditions in imaginary-time. 
Usually, the projections are used to calculate the second derivative
of the energy with respect to momentum (effective mass) 
\cite{Basile,Ceperley_two}. In Ref.~\cite{Kornilovitch} the projection 
was applied for the first time to the whole polaron spectrum. In this 
scheme, the ground-state dispersion is measured directly, and {\em all} 
$E_{\bf P}$ are calculated simultaneously. Unfortunately, at non-zero
{\bf P} the weight of the polaron path is no longer positive definite,
and one needs to deal with a sign-problem. It turns out, 
however, that the main idea can be reformulated in a way that does 
not require a {\em division} by the average sign, but only taking its 
{\em logarithm}. While the new formulation does not constitute a complete
elimination of the sign-problem, it is more statistically stable and
extends the parameter domain accessible in practical simulations.
Below we derive the new formula, discuss its properties, and apply
to the physically interesting example of the lattice polaron.

\section{A formula for the ground-state dispersion}
\label{sec:two}

Up to our knowledge, the projection relations required for our
purposes were first derived by Basile \cite{Basile} .
For completeness a derivation is given below.
Let $R$ denote a many-body real-space configuration,
and $R+{\bf r}$ a many-body configuration which is a result
of the parallel transport of $R$ by a vector ${\bf r}$.
(Note that the sum of $R$ and {\bf r} is only symbolic. The
dimensionality of $R$ is equal to the number of degrees of freedom,
i.e., very large or infinite, while the dimensionality
of {\bf r} is the dimensionality of space, i.e. 1, 2, or 3.)
States $| R \rangle$ form a complete orthogonal basis,
${\bf I} = \int dR \, |R\rangle \langle R|$, and
$\langle R | R' \rangle = \delta(R-R')$. A different basis is formed 
by the states 
$|n\rangle$ which are characterized by the definite total momentum 
${\bf P}$. One is interested in the {\em projected} partition 
function $Z_{\bf P}$ which includes only states with the given {\bf P}:
\begin{eqnarray}
Z_{\bf P} & \equiv & \sum_n \langle n | e^{-\beta H} | n \rangle
\delta_{{\bf P},{\bf P}_n}             \nonumber \\ 
          &   =    & 
\int dR \, dR' \langle R' | e^{-\beta H} | R \rangle \cdot Q_{\bf P} , 
\label{one}
\end{eqnarray}
\begin{equation}
Q_{\bf P} = \sum_n \langle R | n \rangle \langle n | R' \rangle \,
\delta_{{\bf P},{\bf P}_n} =
\langle R | {\bf P} \rangle \langle {\bf P} | R' \rangle ,
\label{two}
\end{equation}
where $\beta=(k_B T)^{-1}$ is the inverse temperature and $H$ is the full 
Hamiltonian. The meaning of Eq.~(\ref{two}) is 
that the two configurations, $R$ and $R'$, have to be projected on the 
given momentum ${\bf P}$. This is achieved as follows (below $\hbar=1$
is set). Any arbitrary configuration $R$ generates a set of states
$|{\bf P}_R \rangle = V^{-1/2} \int d{\bf r} e^{-i {\bf P r}} 
|R+{\bf r} \rangle$, where $V$ is the volume. Inversely, 
$|R+{\bf r}\rangle = V^{-1/2} \sum_{\bf P} e^{i {\bf P r}}
|{\bf P}_R \rangle$. Upon projection, only {\bf P}-components
of both configurations survive. As a result
\begin{eqnarray}
Q_{\bf P} & = & \frac{1}{V} \int d{\bf r}  
\langle R+{\bf r} | {\bf P} \rangle \langle {\bf P} | R'+{\bf r} \rangle
                                                  \nonumber \\  
  & = & \frac{1}{V^2} \int d{\bf r} \sum_{\bf P' P''}
e^{i ({\bf P'-P''}) {\bf r}} 
\langle {\bf P'}_R | {\bf P} \rangle \langle {\bf P} | {\bf P''}_{R'} \rangle
                                                  \nonumber \\
  & = & \frac{1}{V} 
\langle {\bf P}_R | {\bf P}_{R'} \rangle          \nonumber \\
  & = & \frac{1}{V^2} \int d{\bf r} \, d{\bf r'} \, 
\langle R + {\bf r} | R' + {\bf r'} \rangle \,
e^{i {\bf P} ({\bf r}-{\bf r'})}                  \nonumber \\ 
  & = & \frac{1}{V}
\int d(\triangle {\bf r}) \langle R + \triangle {\bf r} | R' \rangle \,
e^{i {\bf P} \triangle {\bf r}}                   \nonumber \\ 
  & = & \frac{1}{V} \int d(\triangle {\bf r}) \, 
e^{i {\bf P} \triangle {\bf r}} \delta((R+\triangle{\bf r}) - R') ,
\label{three}
\end{eqnarray}
where $\triangle {\bf r} = {\bf r} - {\bf r}'$.
Substitution in Eq.~(\ref{one}) and integration over $R'$ yields
\begin{eqnarray}
Z_{\bf P} & = & 
\frac{1}{V} \int \! d(\triangle {\bf r}) \, 
e^{i {\bf P} \triangle {\bf r}} \! \int \! dR 
\langle R + \triangle {\bf r} | e^{-\beta H} | R \rangle \nonumber \\ 
          & = & 
\frac{1}{V}
\int \!d(\triangle {\bf r}) \, e^{i {\bf P} \triangle {\bf r}} \!
\int \! dR \rho(R,R+\triangle{\bf r};\beta) , 
\label{four}
\end{eqnarray}
where $\rho(R,R';\beta)$ is the full many-body density matrix. Next, 
we assume that for each {\bf P} the state with the lowest energy
$E_{\bf P}$ is non-degenerate, and in the low-temperature limit
the projected partition function is dominated by the contribution
from this state, $Z_{\bf P} \rightarrow \exp(-\beta E_{\bf P})$.
Now take the {\em ratio} of $Z_{\bf P}$ and $Z_{{\bf P}=0}$:
\begin{eqnarray}
e^{-\beta(E_{\bf P}-E_0)} & = & \lim_{\beta \rightarrow \infty}
\frac{Z_{\bf P}}{Z_{{\bf P}=0}}  
\label{five}               \\
& = &  \lim_{\beta \rightarrow \infty}
\frac{\int \!d(\triangle {\bf r}) \, e^{i {\bf P} \triangle {\bf r}} \!
\int \! dR \rho(R,R+\triangle{\bf r};\beta) }
{\int \!d(\triangle {\bf r}) \, 
\int \! dR \rho(R,R+\triangle{\bf r};\beta) } , \nonumber
%\label{five}
\end{eqnarray}
where $E_0$ is the ground-state energy. The rhs is nothing but the
average value of $\cos {\bf P} \triangle {\bf r}$ taken over
the distribution $\rho$. [We have assumed that 
$\int dR \rho(R,R+\triangle {\bf r}; \beta)$ is an even function of 
$\triangle {\bf r}$.] A simple formula for $E_{\bf P}$ now follows
\begin{equation}
E_{\bf P} - E_0 = - \lim_{\beta \rightarrow \infty}
\frac{1}{\beta} \ln \langle \cos {\bf P} \triangle {\bf r} \rangle ,
\label{six}
\end{equation}
which is the main result of this section.

Eq.~(\ref{six}) shows that the ground-state dispersion can 
be obtained from the end-to-end distribution of many-body paths. 
It offers a direct way of evaluating the ground-state dispersion
by QMC methods in cases when $\rho(R,R';\beta)$ is positive-definite.
However, the QMC process
must be organized in a special way, as apparent from Eq.~(\ref{five}).
It must generate only such paths whose end configurations at
imaginary time $\tau=\beta$ are exact images of the end configurations
at $\tau=0$ {\em except} for a parallel transport by an arbitrary
vector $\triangle {\bf r}$. It is allowed to change $\triangle {\bf r}$,
make simultaneous changes of both end configurations, and to
make arbitrary changes of paths at internal times $0< \tau < \beta$,
but the end configurations must always be kept identical up to a shift.
It is important that this restriction affects neither the ergodicity 
nor the applicability of the Metropolis algorithm.

Formula (\ref{six}) involves only one measures quantity, 
$\langle \cos {\bf P}\triangle {\bf r} \rangle$, instead of two
in the previous formulation. Moreover, it does not require
division by the measured quantity, but only taking its logarithm.
Additionally, Eq.~(\ref{six}) provides the difference between
two large numbers, $E_{\bf P}$ and $E_0$, and a large cancellation of
errors may occur. This makes Eq.~(\ref{six}) much more stable
statistically than explicit energy estimators.

On the other hand, at small temperatures, the average cosine becomes
exponentially small and it cannot be measured reliably. This
reflects the fact that configurations with $E_{\bf P}-E_0 \gg k_BT$
are very rare because of the Boltzmann factor. Thus, the present method
is limited to excitation energies of the order of several $k_BT$.

\section{Application to the lattice polaron}
\label{sec:three}

We now demonstrate the practical importance of Eq.~(\ref{six}) on
the model problem of lattice polaron, which is often considered as
a paradigmatic example of a particle strongly interacting with
a boson field. We consider a hypercubic lattice with the 
nearest-neighbor hopping, dispersionless phonons, and the
``density-displacement'' electron-phonon interaction. 
The model Hamiltonian reads
\begin{equation}
H = - t \sum_{\langle {\bf nn'} \rangle} c^{\dagger}_{\bf n} c_{\bf n'}
- \sum_{\bf nm} f_{\bf m}({\bf n}) c^{\dagger}_{\bf n} c_{\bf n} \xi_{\bf m}
+ \hbar\omega \sum_{\bf m} b^{\dagger}_{\bf m} b_{\bf m} .
\label{seven}
\end{equation}
Here $t$ is the hopping amplitude (it will be used as the energy unit),
$\omega$ is the phonon (oscillator) frequency, $\xi_{\bf m}$ is the internal
coordinate of the {\bf m}th oscillator, and $f_{\bf m}({\bf n})$ is
the force between {\bf m}th oscillator and the particle at site {\bf n}
($f$ is a function of distance $|{\bf m-n}|$ only). The model is
parametrized by the dimensionless frequency $\bar\omega=\hbar\omega/t$
and by the dimensionless coupling constant 
$\lambda=[\sum_{\bf m} f^2_{\bf m}(0)]/(2M\omega^2\,D)$, where
$M$ is the mass of the oscillator and $D$ is the half-bandwidth
of the bare band. (For an isotropic band with nearest-neighbor
hopping, $D = zt$, $z$ being the number of neighbors.) 

For the polaron problem, a many-body configuration $R$ is specified
by the position of the electron {\bf r} and oscillator displacements
$\xi{\bf m}$. Making use of the Feynman's idea of analytic integration
over $\xi{\bf m}$ \cite{Feynman} the problem is reduced to a 
single-particle system with retarded self-interaction. The latter
can be simulated exactly, using the continuous-time representation
of polaron paths \cite{Beard}. The resulting algorithm \cite{Kornilovitch}
is very efficient and allow accurate determination of the ground-state
energy and effective mass of the polaron for a wide class of models.
It this paper, it will be shown that the method also produces accurate
polaron spectra, when combined with Eq.~(\ref{six}). There are other
reasons why the polaron is an ideal system to try formula (\ref{six}).
First, due to a constant phonon frequency, excited states are, at any 
{\bf P}, separated from the restricted ground state by a finite energy 
gap $\hbar\omega$. Therefore, instead of performing numerically the 
limit procedure to $\beta = \infty$, one can study the system at 
{\em finite} $\beta$, provided $\exp(\beta \hbar\omega) \gg 1$ and 
the contribution from excited states is negligible. Second, by increasing 
the coupling constant $\lambda$ one can always decrease $E_{\bf P}-E_0$, 
i.e., substantially increase 
$\langle \cos {\bf P} \triangle {\bf r} \rangle$, and stabilise 
the sumulations. Third, the polaron momentum {\bf P} is not
a parameter of simulations. This implies that statistics can be 
collected for all momenta simultaneously. In other words, the whole
polaron spectrum is measured in a single QMC run. This will enable
us to calculate for the first time exact polaron densities of states.

\begin{figure}[ht]
\begin{center}
\leavevmode
\hbox{
\epsfxsize=8.2cm
\epsffile{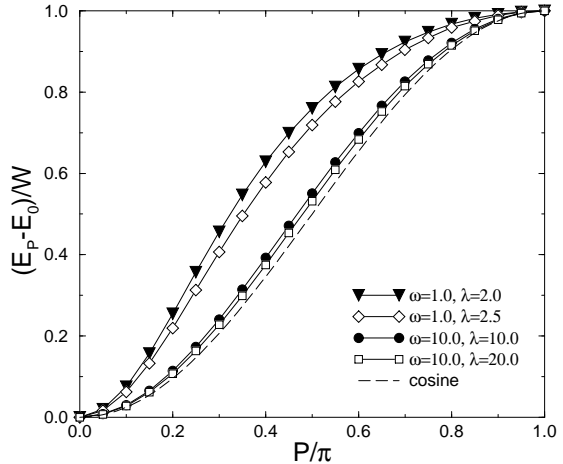}
}
\end{center}
\caption{
Polaron spectrum in the one-dimensional Holstein model, normalized
to the bandwidth $W = E_{\pi}-E_0$. Triangles: $\bar\omega=1.0$,
$\lambda=2.0$ [for these parameters the polaron ground-state energy
$E_0=-4.38(1)$ (in units of $t$), bandwidth $W=0.1243(2)$ (in units 
of $t$), effective mass $m^{\ast}=10.0(1)$ (in units of 
$m_0=\hbar^2/(2ta^2)$).] 
Diamonds: $\bar\omega= 1.0$, $\lambda= 2.5$ 
[$E_0=-5.26(1)$, $W=0.0437(3)$, $m^{\ast}=34.5(3)$] .
Circles:  $\bar\omega=10.0$, $\lambda=10.0$ 
[$E_0=-20.35(1)$, $W=0.543(2)$, $m^{\ast}=6.06(2)$] .
Squares:  $\bar\omega=10.0$, $\lambda=20.0$ 
[$E_0=-40.08(1)$, $W=0.0739(2)$, $m^{\ast}=47.6(1)$] .
}
\label{fig1}
\end{figure}

We begin with the simplest Holstein model which has local
electron-phonon interaction, 
$f_{\bf m}({\bf n}) = \kappa \delta_{\bf mn}$. 
In one dimension, the polaron spectrum has been extensively studied by 
exact diagonalisation \cite{Fehskeone,Fehsketwo},
strong-coupling perturbation \cite{Stephan}, and variational 
\cite{Brown} techniques. Our QMC data for the one-dimensional 
Holstein model are shown in Fig.~\ref{fig1}. 
The most interesting feature of the spectrum is its non-cosine
shape in the adiabatic regime $\bar\omega \leq 1.0$ (triangles and
diamonds). At large momenta, the spectrum is more flat than at small 
momenta. The nature of this flattening was understood a long time 
ago \cite{Rashba}. In the weak-coupling limit,
the free-particle state hybridizes with the single-phonon state 
and creates a mixed ground state which is free-particle-like
at small P and phonon-like at large P, hence the weak
dispersion. With increasing $\lambda$, the free-particle state
is replaced with a polaron state with an effective mass $m^{\ast}$,
which now hybridizes with the single-phonon state, still leading
to a more flat dispersion at large momenta.
The flattening effect weakens with growing $\lambda$ and $\bar\omega$
because both processes increase the energy separation of the two 
hybridizing states. In recent years the flattening of the polaron
spectrum was observed in numerical studies \cite{Fehskeone,Stephan}. 

Our Quantum Monte Carlo data fully confirm the previous analytical 
and numerical results, see Fig.~\ref{fig1}. We found that the 
spectrum shape is more sensitive to the phonon frequency than to 
the coupling constant. At a small frequency $\bar\omega=1.0$, 
the increase 
of the coupling constant from $\lambda=2.0$ to $\lambda=2.5$ results
in a 3.5-times increase of the effective mass, and in a 2.8-times
drop of the bandwidth, yet the spectrum shape changes only slightly,
cf. triangles and diamonds in Fig.~\ref{fig1}. At the same time,
a simultaneous 10-times increase of $\lambda$ and $\bar\omega$ results
in a similar, 4.8 times, increase of effective mass but brings the 
spectrum shape very close to
cosine, cf. triangles and squares in Fig.~\ref{fig1}. Again, a 
doubling of $\lambda$ strongly affects $E_0$ and $W$ but not the
spectrum shape, cf. circles and squares in Fig.~\ref{fig1}.  

\begin{figure}[ht]
\begin{center}
\leavevmode
\hbox{
\epsfxsize=8.2cm
\epsffile{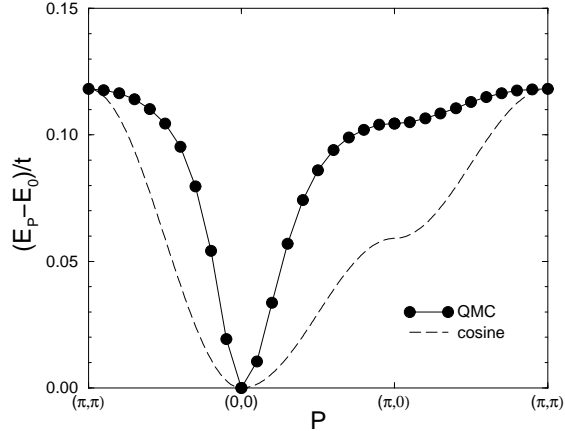}
}
\end{center}
\caption{
Spectrum of the two-dimensional Holstein model in the adiabatic regime.
$\bar \omega = 1.0$, $\lambda = 1.4$. 
[$E_0=-6.12(3)$, $m^{\ast}=8.7(1)$.]
}
\label{fig2}
\end{figure}

It is instructive to compare the exact QMC results with the Lang-Firsov (LF)
approximation \cite{Lang} which is believed to be the correct description
of the polaron in the antiadiabatic regime (high phonon frequency).
The LF formula for the spectrum reads
\begin{equation}
E_P - E_0 = 2t \, e^{-z\lambda/\bar\omega} ( 1 - \cos{P}), 
\label{nine}
\end{equation}
which also implies the relation between the renormalized mass and 
bandwidth
\begin{equation}
\frac{m^{\ast}}{m_0} = \frac{W_0}{W} = e^{z\lambda/\bar\omega} ,
\label{nineandhalf}
\end{equation}
where $W_0=2zt$ is the bare bandwidth and $m_0 = \hbar^2/(2ta^2)$ is 
the bare mass, $a$ being the lattice constant. For 
$\bar\omega=10.0$ and $\lambda=10.0$ QMC results are $W=0.543(2)\,t$ 
and $m^{\ast}=6.06(2)\,m_0$ while LF yields $W_{LF}=0.541\,t$ and 
$m^{\ast}_{LF}=7.39\,m_0$. For $\bar\omega=10.0$ and 
$\lambda=20.0$ one has $W=0.0739(2)\,t$ and $m^{\ast}=47.6(1)\,m_0$ 
from QMC and $W_{LF}=0.0733\,t$ and $m^{\ast}_{LF}=54.6\,m_0$ from LF.
One can see that LF predicts very accurate values of the polaron
bandwidth. This fact was established in the previous studies of the
Holstein model \cite{Kabanov,Fehskeone}. On the other hand, LF
slightly overestimates the polaron mass. This is due to small
deviations from the cosine shape, still present in the true
spectrum at these model parameters. Still, the LF masses are reasonably
close to the exact ones, and the agreement improves with the further
increase of $\bar\omega$ and $\lambda$.
   
Consider now the two-dimensional Holstein model. The only exact 
polaron spectra published so far were calculated by 
Wellein, Fehske, and Loos with the exact diagonalization method 
\cite{Fehsketwo}. These authors found a flattening of the spectrum in
the outer part of the Brillouin zone, even stronger than in the 
one-dimensional case. We checked that for the model parameters 
used in \cite{Fehsketwo}, formula (\ref{six}) yields precisely the 
same values of $E_{\bf P}-E_0$ as the exact diagonalization method. 
A definite advantage of the present method is that it allows 
simultaneous calculations at any desired number of {\bf P}-points, 
while exact diagonalization studies are limited to a small number of
{\bf P}-points due to the finite size of the clusters.
On the other hand, the QMC method is limited to the condition
$W \ll \hbar\omega$, which prevents us from studying the 
weak-coupling regime and such an interesting phenomenon as
the limit point of the polaron spectrum \cite{Rashba}. (The latter
is possible with the diagrammatic QMC \cite{Prokof'ev}.)
Figure~\ref{fig2} shows our QMC data for a new set of parameters
in the adiabatic regime,
$\bar \omega = 1.0$, $\lambda = 1.4$, where 30 {\bf P}-points
have been used to represent the spectrum. One can see that
the dispersion is indeed weak for $|{\bf P}| > \pi/2$, i.e. in
the {\em larger} part of the Brillouin zone. Note, that at
these parameters the polaron bandwidth is reduced by 
$8\,t/0.12\,t = 67$ times, and the mass enhancement is 8.7,
so we are already in the small polaron regime. Yet, the
spectrum shape is profoundly non-cosine. With increasing
$\lambda$, it will be approaching the cosine shape, but this
is expected to happen only at such large $\lambda$ where 
the polaron is very heavy and is easily localized.  

\begin{figure}[ht]
\begin{center}
\leavevmode
\hbox{
\epsfxsize=8.2cm
\epsffile{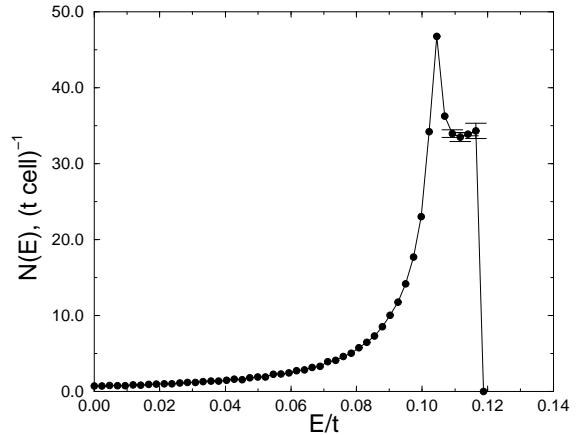}
}
\end{center}
\caption{
Density of states of the two-dimensional Holstein model in the
adiabatic regime. $\bar \omega = 1.0$, $\lambda = 1.4$.
[$E_0=-6.12(3)$, $m^{\ast}=8.7(1)$.]
}
\label{fig3}
\end{figure}

In two dimensions, the volume of the outer part of the Brillouin zone
is larger than that of the inner one. Therefore the linear
representation of the spectrum, like in Fig.~\ref{fig2}, does
not fully convey the changes in the band structure, caused by
the flattening of the spectrum. The proper physical quantity
which takes into account all the states of the Brillouin zone 
is the density of states (DOS). The QMC method, 
coupled with formula (\ref{six}), provides the unique opportunity
to calculate polaron DOS exactly, since it allows a simultaneous
measurement of the whole spectrum.
In this work, the two-dimensional Brillouin zone was divided in 
$200^2$ points at which the spectrum was measured. 
In the end, the total of $40\,000$ polaron energies were distributed 
over 50 energy intervals between 0 and $W$. The resulting DOS for
$\bar \omega = 1.0$, $\lambda = 1.4$ (the same parameters as in
Fig.~\ref{fig2}) is shown in Fig.~\ref{fig3}. One can see that
the effect of the spectrum flattening is indeed quite dramatic.
The upper half-band is jammed into a narrow, $0.015\,t$-width, 
energy interval, thereby increasing DOS at the top of the band
to $\approx 50$ times the DOS at the bottom of the band. The
van Hove singularity is shifted from the middle to the top of the
band. The lower half of the band contains only $13\%$ of all states.
Overall, DOS looks qualitatively different from the free-particle
one.

\begin{figure}[ht]
\begin{center}
\leavevmode
\hbox{
\epsfxsize=8.2cm
\epsffile{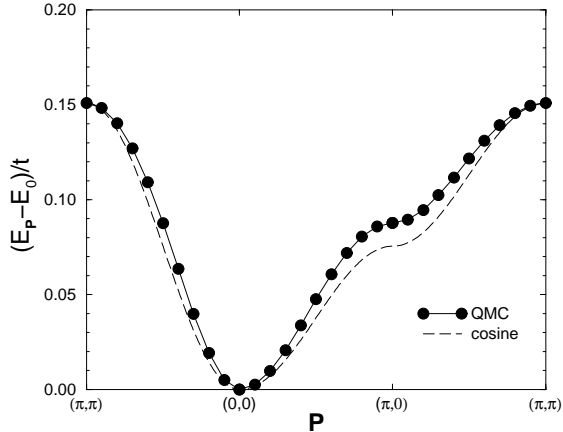}
}
\end{center}
\caption{
Spectrum of the two-dimensional Holstein model in the anti-adiabatic
regime. $\bar \omega = 8.0$, $\lambda = 8.0$.
[$E_0=-32.16(1)$, $m^{\ast}=38.4(1)$.]
}
\label{fig4}
\end{figure}
\begin{figure}[ht]
\begin{center}
\leavevmode
\hbox{
\epsfxsize=8.2cm
\epsffile{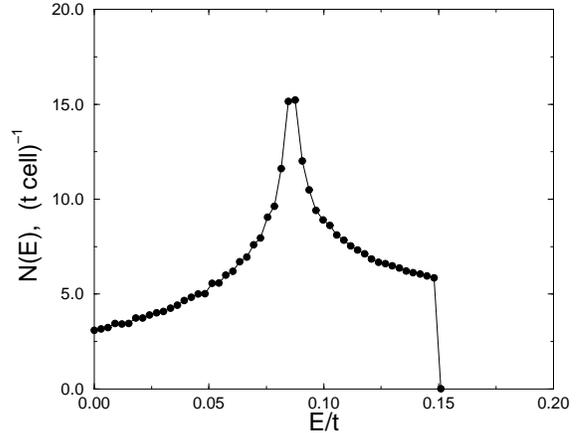}
}
\end{center}
\caption{
Density of states of the two-dimensional Holstein model in the
anti-adiabatic regime. $\bar \omega = 8.0$, $\lambda = 8.0$.
[$E_0=-32.16(1)$, $m^{\ast}=38.4(1)$.]
Small fluctuations of DOS are artifacts of a finite-mesh 
integration over the Brillouin zone and not statistical errors.
}
\label{fig5}
\end{figure}

As in the one-dimensional case, the band structure approaches
the free-particle-like as the phonon frequency increases. As an
example, we considered a frequency equal to the bare bandwidth,
$\bar\omega=8.0$, and $\lambda=8.0$. (This value of the coupling
constant was chosen to have the polaron bandwidth, $W$, close to 
the previously considered adiabatic case.) The polaron spectrum and
DOS are shown in Figs.~\ref{fig4} and \ref{fig5}, respectively.
Deviations from the free-particle behavior are still visible, but
they are small and not qualitative. DOS at the top of the band is just 
$\approx 1.5$ times larger than at the bottom, and the singularity
appears close to the middle of the band. The ``normalization''
of the spectrum at large $\bar\omega$ is quite understandable.
With increasing phonon frequency, the retardation effects become
less important, the lattice deformation more readily follows
the particle movement, and the whole complex behaves more 
like a free particle with a renormalized hopping integral.
The Lang-Firsov formula (\ref{nineandhalf}) predicts $W_{LF}=0.1465\,t$
and $m^{\ast}_{LF}=54.6\,m_0$ which is to be compared with the QMC
results $W=0.1510(3)\,t$ and $m^{\ast} = 38.4(1)\,m_0$. Again,
the LF approximation yields the correct bandwidth but overestimates
the effective mass by some 40\%.

\begin{figure}[ht]
\begin{center}
\leavevmode
\hbox{
\epsfxsize=8.2cm
\epsffile{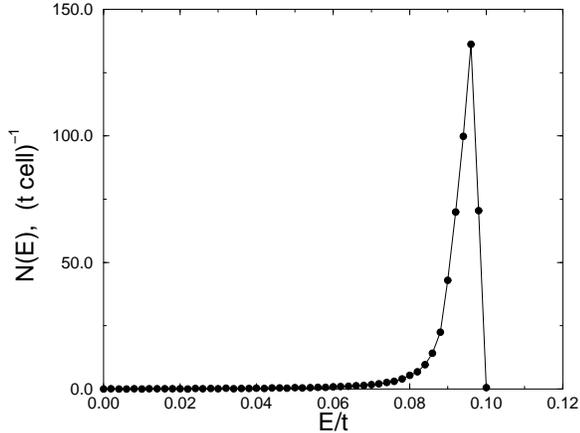}
}
\end{center}
\caption{
Density of states of the three-dimensional Holstein model in the
adiabatic regime. $\bar \omega = 1.0$, $\lambda = 1.2$.
[$E_0=-7.75(4)$, $m^{\ast}=6.2(2)$.]
}
\label{fig6}
\end{figure}
\begin{figure}[ht]
\begin{center}
\leavevmode
\hbox{
\epsfxsize=8.2cm
\epsffile{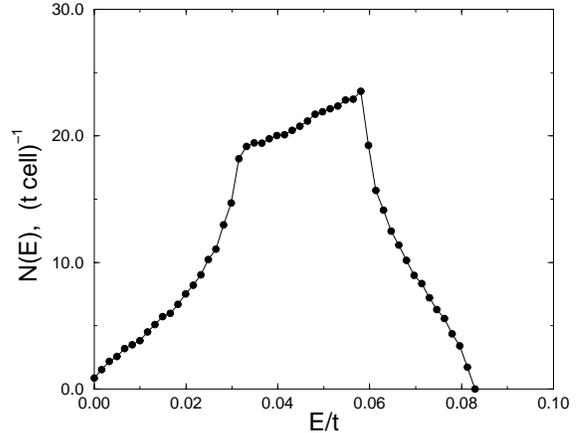}
}
\end{center}
\caption{
Density of states of the three-dimensional Holstein model in the
anti-adiabatic regime. $\bar \omega = 12.0$, $\lambda = 10.0$.
[$E_0=-60.12(2)$, $m^{\ast}=112(1)$.]
Small fluctuations of DOS are artifacts of a finite-mesh 
integration over the Brillouin zone and not statistical errors.
}
\label{fig7}
\end{figure}

The most spectacular transformation of DOS occurs in the three-dimensional
Holstein model. In three dimensions, the volume of the outer part 
of the Brillouin zone is {\em much} larger than that of the inner 
part, and it should completely dominate the total DOS. We do not 
show the polaron spectrum which is not very informative. 
Density of states was calculated by measuring the 
polaron spectrum at $60^3$ points of the full Brillouin zone, and 
then distributing them among 50 energy intervals between 0 and $W$. 
DOS in the adiabatic regime, $\bar\omega=1.0$ and $\lambda=1.2$, is 
shown in Fig.~\ref{fig6}. The states of the flat part of the spectrum 
form a massive peak at the top of the band. The width of the peak is 
about $10\%$ of the total bandwidth. DOS at the bottom of the band is 
negligible, the capacity of the lower half-band is less than $1\%$ 
of the total number of states. The two van Hove singularities are 
not visible, at least on the chosen level of energy resolution, 
they are absorbed into the peak. Overall, DOS is 
{\em completely} different from the free-particle one.
Should the three-dimensional Holstein model with such parameters
exist in nature, an extreme care would be necessary in interpreting
experimental data. In any real material, the lowest states would
likely be localized, and any response to an external perturbation
would be dominated by the peak. Then, for instance, fitting to a 
free-particle-like form of DOS would lead to wrong estimates of the 
coupling constant and other errors. 

As before, the band structure returns to the free-particle shape 
in the anti-adiabatic regime. We considered the case of the phonon 
frequency being equal to the bare
bandwidth, $\bar\omega=12.0$, and coupling constant $\lambda=10.0$,
when the polaron bandwidth is close to the just considered adiabatic
case, see Fig.~\ref{fig7}. Although still distorted, the DOS shape
is close to the free-particle one, with the square-root behavior
at the top and the bottom of the band, and with two van Hove 
singularities fully developed at the ``right'' places. The polaron
bandwidth is $W=0.0827(2)\,t$, which is in good agreement with the 
Lang-Firsov value $W_{LF}=0.0809\,t$, while the polaron mass 
$m^{\ast} = 112(1)\,m_0$ is 24\% lighter than the LF mass 
$m^{\ast}_{LF}=148\,m_0$.

\begin{figure}[ht]
\begin{center}
\leavevmode
\hbox{
\epsfxsize=8.2cm
\epsffile{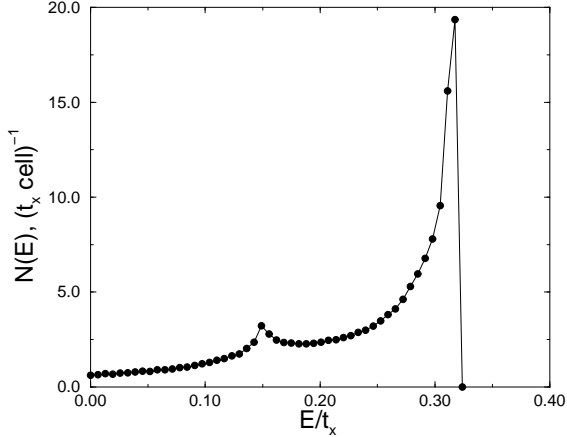}
}
\end{center}
\caption{
Density of states of the two-dimensional anisotropic Holstein model.
$\bar \omega = 1.0$, $\lambda = 1.4$, $t_y/t_x = 0.2$.
[$E_0=-3.987(3)$, $m^{\ast}_x=3.44(1)\,m_{0x}$, 
$m^{\ast}_y=17.54(3)\,m_{0x}$.]
}
\label{fig8}
\end{figure}

The polaron QMC algorithm of Ref.~\cite{Kornilovitch} is not
limited to the Holstein model. In fact, it allows studies of
{\em arbitrary} forms of the electron-phonon interactions
(of the density-displacement type), and arbitrary forms of
the particle kinetic energy. Combined with Eq.~(\ref{six}),
it provides an efficient and exact way of calculating the band 
structure of a whole class of polaron models. As possibilities
are numerous, we have chosen to illustrate the point on two
particular examples. 

The first example is the {\em anisotropic} two-dimensional Holstein
model with $\bar\omega=1.0$, $\lambda=1.4$ (these parameters are
the same as in Figs.~\ref{fig2} and \ref{fig3}), and the bare
anisotropy ratio $t_y/t_x=0.2$. For a free particle with such an
anisotropy, the saddle points at $(\pm\pi,0)$ and $(0,\pm\pi)$ 
have different energies, which results in two singularities
in DOS, positioned symmetrically with respect to the center and
edges of the band. Polaron DOS, calculated by QMC, is shown in 
Fig.~\ref{fig8}. The flattening effect creates a strong peak
at the top of the band which absorbs the higher-energy singularity
[at $(\pm\pi,0)$]. At the same time, the second singularity
is still clearly visible. Now it appears in the vicinity of
the middle of the band.    

\begin{figure}[ht]
\begin{center}
\leavevmode
\hbox{
\epsfxsize=8.2cm
\epsffile{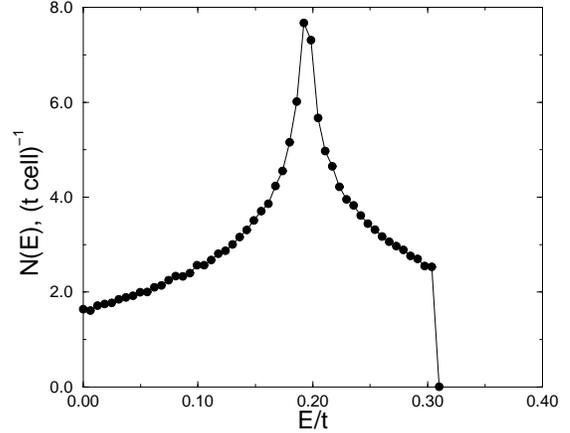}
}
\end{center}
\caption{
Density of states of the two-dimensional model with long-range
electron-phonon interaction. $\bar \omega = 1.0$, $\lambda = 2.75$.
[$E_0 = -11.83(1)$, $m^{\ast} = 20.1(1)$.]
Small fluctuations of DOS are artifacts of a finite-mesh 
integration over the Brillouin zone and not statistical errors.
}
\label{fig9}
\end{figure}

The second example is the two-dimensional polaron model with
{\em long-range} electron-phonon interaction [combine with the
Hamiltonian (\ref{seven})]:
\begin{equation}
f_{\bf m}({\bf n}) = \frac{\kappa}
{(| {\bf m}-{\bf n} |^2 + 1)^{3/2}} ,
\label{eight}
\end{equation}
where the distance $| {\bf m}-{\bf n} |$ is measured in lattice
constants. For this form of the force, 
$\lambda=1.742 \kappa^2/(2M\omega^2 D)$.
The model describes a two-dimensional particle interacting with 
a {\em parallel} plane of ions vibrating perpendicular to the 
plane. It was proposed in Ref.\cite{Alexandrov},
where it was used to model the interaction of holes doped in 
copper-oxygen planes, with apical oxygens in the layered cuprates. 
It was found that this long-range (Fr\"ohlich) polaron is much 
lighter than the short-range Holstein polaron.   
Here we present the density of states for $\bar\omega=1.0$ and
$\lambda=2.75$, see Fig.~\ref{fig9}. DOS shape is close to the
free-particle one, with a single, well-developed singularity
in the middle of the band. Note, that we are in the adiabatic
regime, at the same frequency $\bar\omega=1.0$ where the two-dimensional
Holstein polaron has a very distorted DOS, cf. Fig.~\ref{fig3}.
The comparison of Figs.~\ref{fig9} and \ref{fig5} shows that
a long-range electron-phonon interaction plays the same role
as increasing phonon frequency, as far as the flattening effect
is concerned.

\section{Discussion and conclusions}
\label{sec:four}

The main message of this paper is that a path-integral imaginary-time 
Quantum Monte Carlo is quite capable of direct measuring of 
{\em real-time} spectra. By relaxing the boundary conditions in 
imaginary time, one allows many-body paths to have arbitrary 
real-space shift $\triangle {\bf r}$. Then the Fourier transform
projects out configurations with a certain total momentum {\bf P}.
The result, expressed by the formula (\ref{six}), is  
the ground-state energy as a function of the 
total momentum. The physical system, where such a ground-state 
dispersion is of interest, should be carefully chosen. 
 
The role of temperature in this process is two-fold.
On one hand, temperature should be made as low
as possible to exclude the contribution from the excited
states with the same {\bf P}. On the other hand, only states
with $E_{\bf P}-E_0 \sim k_B T$ are excited in the system. This means
that the corresponding configurations will be generated by
the QMC process in amounts sufficient for a good statistics. 
For higher-energy states, configurations will 
be exponentially rare. This is the reason for the average
cosine in Eq.~(\ref{six}) to become exponentially small in the
low-temperature limit. In this 
case, the measurement process will be statistically unstable.    
Thus, the temperature should be of the order of the energy
interval one is interested in. The two conditions on the temperature
can be reconciled if the energy scale of the ground-state dispersion
is much smaller than the energy gap between the ground state and
the excited states within the same {\bf P}-sector. This is realized
in the polaron system where one has $W \ll \hbar\omega$ 
for a wide range of parameters. If this condition is not satisfied,
Eq.~(\ref{six}) will be measuring the difference of projected
free energies rather than that of ground-state energies. 

Eq.~(\ref{six}) shows that in a many-body system there exists a 
general and simple relation between the ground-state 
dispersion and the end-to-end distribution of imaginary-time paths. 
It does not involve the energy estimator, although the latter
is required for the separate calculation of $E_0$ and $E_{\bf P}$.
This might be useful in cases when the evaluation of the energy
estimator is computationally costly. There are other
computational advantages. First, Eq.~(\ref{six}) involves only
the logarithm of one measured quantity, the average cosine.
Second, Eq.~(\ref{six}) calculates the difference of two energies 
both of which may be large. In the polaron problem, typical energies 
are of the order of a few $t$ but the bandwidth is $W \sim 0.1 \,t$. 
Both $E_0$ and $E_{\bf P}$ can be calculated with typical accuracy 
$0.3-0.5\,\%$. This may result in a sizable error in their difference 
if the two energies are calculated separately and then subtracted. 
Eq.~(\ref{six}) produces much more stable energy differences because
of the large cancellation of errors between $E_{\bf P}$ and $E_0$.
In all the spectra
presented in this paper, Figs.~\ref{fig1}, \ref{fig2}, \ref{fig4},
the statistical errors are smaller that symbols representing the data.
Finally, the whole dispersion (as well as $E_0$, and derivatives at
any {\bf P}-point) can be calculated during a {\em single} QMC run.
This property of Eq.~(\ref{six}) also allows fast computation of the
density of states.

To demonstrate the practical usefulness of Eq.~(\ref{six}), 
we have combined it with an exact continuous-time
algorithm for the lattice polaron \cite{Kornilovitch}, and
calculated first detailed polaron spectra and densities of
states in two and three dimensions. Although our method of
calculating the spectrum is limited to the condition 
$W \ll \hbar\omega$, i.e., to the intermediate and strong
coupling, this is the most physically interesting regime. 
Together with the weak- and strong-coupling perturbation expansions, 
the exact diagonalization, density-matrix renormalization group 
\cite{White}, variational, and diagrammatic QMC techniques,  
the present method covers the {\em whole} parameter
range of the polaron problems. One can say, that the problem
of calculating the exact polaron spectrum has found its solution.

Apart from being a test for Eq.~(\ref{six}), the
lattice polaron still has a considerable interest on its own.
We have seen in this paper that the polaron spectrum in the
adiabatic limit of the Holstein model is generically non-cosine,
as was predicted theoretically and recently observed numerically.
The flattening has been found to continue well into the 
strong-coupling regime, where the polaron mass is a few dozens
and approaches a hundred. The same conclusion was reached 
previously in \cite{Fehsketwo}. For physical applications the most
interesting regime is when polarons are not very heavy and can 
be mobile. We conclude that {\em the non-cosine spectrum is typical
for the Holstein model in the physically relevant parameter region},
i.e., small phonon frequencies and intermediate couplings.

A surprise finding has been the great extent at which the flattening 
changes the band structure in high dimensions. Density of states
is completely changed, van Hove singularities are shifted, 
low-lying states are almost irrelevant, relations between the
effective mass and the bandwidth is broken. We have seen that the
combination of dimensionality, phonon frequency, coupling strength,
and anisotropy may produce densities of states of various shapes.
In such a situation, one should be careful about the
interpretation of any experimental data, like the estimation 
of $\lambda$ or polaron mass from the bandwidth or from the 
location of singularities. The use of a ``naive'' model band 
structure would lead to wrong conclusions.  

We have also found that in all dimensions the polaron band structure 
becomes free-particle-like with increasing phonon frequency. 
In particular, the spectrum approaches the cosine shape and the 
effective mass approaches the inverse of the bandwidth. Moreover, 
numerical values of $W$ are well described by the Lang-Firsov 
approximation. Thus, our QMC data support the LF approximation 
as the right description of the polaron in the antiadiabatic 
regime. At the same time, we have found that LF could still 
overestimate the polaron mass by a few dozen percent in cases
where the bandwidth is predicted correctly. 

Finally, we considered a long-range electron-phonon interaction
and found a free-particle-like band structure even in the adiabatic
regime. In Ref.~\cite{Alexandrov} it was found that in the adiabatic 
regime polaron masses are well described by the LF approximation. 
Two conclusions follow from these facts. First, all the unusual
properties of the Holstein model caused by the flattening,
may be specific to the local electron-phonon interaction and
may not be generic polaron properties. Second, the long-range
electron-phonon interaction on a lattice seems to have the
same effect on the band structure as the increasing phonon frequency.
Although it is clear intuitively that a long-range interaction
leads to higher mobility of the lattice deformation, details of
this mechanism are yet to be fully understood.

\vspace{0.5cm}
The author is grateful to A.\,S.\,Alexandrov, D.\,M.\,Ceperley, V.\,Elser, 
W.\,M.\,C.\,Foulkes, and V.\,V.\,Kabanov for useful discussions
and communications. This work was supported by EPSRC under grant GR/L40113.


\begin{references}


\bibitem{Ceperley}
E.\,L.\,Pollock and D.\,M.\,Ceperley,
Phys.\,Rev.\,B {\bf 36}, 8343 (1987).

\bibitem{Scalapino}
D.\,J.\,Scalapino, S.\,R.\,White, and S.\,C.\,Zhang,
Phys.\,Rev.\,B {\bf 47}, 7995 (1993).

\bibitem{Basile}
A.\,G.\,Basile, 
PhD thesis, (Cornell University, 1992).

\bibitem{Ceperley_two}
M.\,Boninsegni and D.\,M.\,Ceperley,
Phys.\,Rev.\,Lett. {\bf 74}, 2288 (1995). 

\bibitem{Alexandrou}
C.\,Alexandrou and R.\,Rosenfelder,
Phys.\,Rep. {\bf 215}, 1 (1992).

\bibitem{Kornilovitch}
P.\,E.\,Kornilovitch,
Phys.\,Rev.\,Lett. {\bf 81}, 5382 (1998).

\bibitem{Prokof'ev}
N.\,V.\,Prokof'ev and B.\,V.\,Svistunov,
Phys.\,Rev.\,Lett. {\bf 81}, 2514 (1998).

\bibitem{Feynman}
R.\,P.\,Feynman,
Phys.\,Rev. {\bf 97}, 660 (1955);
{\em Statistical Mechanics} (Benjamin, Reading, 1972).

\bibitem{Beard}
B.\,B.\,Beard and U.-J.\,Wiese,
Phys.\,Rev.\,Lett. {\bf 77}, 5130 (1996).

\bibitem{Fehskeone}
G.\,Wellein, H.\,R\"oder, and H.\,Fehske,
Phys.\,Rev.\,B {\bf 53}, 9666 (1996).

\bibitem{Fehsketwo}
G.\,Wellein, and H.\,Fehske,
Phys.\,Rev.\,B {\bf 56}, 4513 (1997).
H.\,Fehske, J.\,Loos, and G.\,Wellein,
Z.\,Phys.\,B {\bf 104}, 619 (1997).

\bibitem{Stephan}
W.\,Stephan,
Phys.\,Rev.\,B {\bf 54}, 8981 (1996).

\bibitem{Brown}
A.\,H.\,Romero, D.\,W.\,Brown, and K.\,Lindenberg,
J.\,Chem.\,Phys. {\bf 109}, 6540 (1998);
cond-mat/9809025.

\bibitem{Rashba}
Y.\,B.\,Levinson and \'E.\,I.\,Rashba,
Rep.\,Prog.\,Phys. {\bf 36}, 1499 (1973).

\bibitem{Lang}
I.\,G.\,Lang and Yu.\,A.\,Firsov, 
Zh.\,Eksp.\,Teor.\,Fiz. {\bf 43}, 1843 (1962)
[ Sov.\,Phys.\,JETP {\bf 16}, 1301 (1963)].

\bibitem{Kabanov}
A.\,S.\,Alexandrov, V.\,V.\,Kabanov, and D.\,K.\,Ray,
Phys.\,Rev.\,B {\bf 49}, 9915 (1994).

\bibitem{Note}
The algorithm of Ref.~\cite{Kornilovitch} is very fast in the 
small-polaron regime. It is interesting that in calculation of DOS, 
most of the CPU time is spent on the ``dull'' evaluation of the 
cosine in Eq.~(\ref{six}), about $10^6$ times at each {\bf P}-point.

\bibitem{Alexandrov}
A.\,S.\,Alexandrov and P.\,E.\,Kornilovitch,
Phys.\,Rev.\,Lett. {\bf 82}, 807 (1999).

\bibitem{White}
E.\,Jeckelmann and S.\,R.\,White,
Phys.\,Rev.\,B {\bf 57}, 6376 (1998).

\end{references}
\end{document}